\documentclass[10pt,letterpaper,twocolumn]{article}

\usepackage{ol2_no_copyright}
\usepackage[draft]{hyperref}
\usepackage{amsmath}
\usepackage{graphicx}

\begin{document}
\twocolumn[ 

\title{Far-off resonance conditional phase-shifter using the ac-Stark shift}
\author{N. A. Proite and D. D. Yavuz$^{*}$}
\address{Department of Physics, University of Wisconsin~--~Madison, \\ 1150 University Ave., Madison, WI, 53706\\
$^*$Corresponding author: yavuz@wisc.edu
}
\begin{abstract}
We propose a simple technique that achieves a conditional phase shift of $\pi$ radians between two weak lasers with energies at the 1000-photon level.  The key idea is to set up a V-system with two far-off resonant lasers by coupling the ground state to two excited electronic states.  The lasers interact through the ac Stark shift of the ground state and thereby acquire a large conditional phase shift. \end{abstract}

 ] 
Interacting low-power laser beams is a subject of considerable attention in nonlinear and quantum optics \cite{ScullyBook}. Nonlinear interactions between weak beams can form optical switches with possible applications in all-optical information processing. Furthermore, if achieved at the single photon level, these interactions can also be used to entangle single photons, which may form the basis of a future photonic quantum computing device. In traditional nonlinear materials, the weakness of optical nonlinearities prohibit observing significant nonlinear effects between weak beams.  Over the last decade, suggestions involving Electromagnetically Induced Transparency (EIT) have generated much enthusiasm in this field \cite{SchmidtImamoglu,HarrisYamamoto,HarrisNonlinearEIT,LukinUltraslow,Xiao,Ottaviani,ZhuSwitch}. Recent experiments have demonstrated optical switching at $\sim$10 photons per atomic cross-section using EIT-based approaches \cite{Harris2003,ZhuSwitch}.  Additionally, switching with optical instabilities has been demonstrated in an atomic vapor at less than one photon per atomic cross-section\cite{GauthierVapor}.

In this Letter, we suggest a far-off resonant technique that achieves a nonlinear phase-shift of $\pi$ radians with low absorption in a free-space geometry (without the use of a cavity). Our phase-shifter requires laser energies at the 1000-photon level and can be configured to have a large bandwidth. Although our technique does not achieve large nonlinear phase shift at the single-photon level in a free space geometry, there are key advantages of our scheme when compared with earlier approaches: 1) Our scheme does not require a strong coupling laser as is required by EIT. As a result, the total energy requirement of our switch is at the 1000-photon level. 2) The bandwidth of our switch can be large and one can work with nanosecond time scale optical pulses. The bandwidth can be increased until the rotating-wave approximation breaks down at the expense of an increased density-length product. 3) For sufficiently large detuned beams, Doppler broadening becomes unimportant and as a result, our scheme is well suited for vapor cells. Due to these advantages, our approach may be particularly useful for constructing ultra-low power, high-bandwidth all-optical switches with possible applications in current fiber-optic networks.

As shown in Fig.~\ref{levels}, we begin with a neutral alkali atomic medium containing a ground state $|1\rangle$ and two excited states, $|2\rangle$ and $|3\rangle$.  A probe beam, $E_p$, and a switch beam, $E_s$, are tuned far-off resonant from the $|1\rangle$--$|2\rangle$ and $|1\rangle$--$|3\rangle$ transitions, respectively\footnote{In general, this phase-shifter scheme is not exclusive to V-systems.  For example, one may tune both beams to the same lower and upper states in a two-level scheme.  Then, the switch beam will ac Stark shift both the lower and upper states.}.  Without the switch beam, the weak probe laser will experience phase accumulation and absorption as determined by the linear susceptibility of the atomic medium. These quantities depend on the probe's frequency detuning from the atomic resonance, $\Delta_p$.  When both the probe and switch propagate together through the medium, the detuning of the probe effectively changes. This is because the switch beam will ac Stark shift the common ground state $|1\rangle$. In the presence of the switch beam, the susceptibility of the medium is modified to give:

\begin{figure}[tb]
\begin{center}
\includegraphics[width=8.3cm]{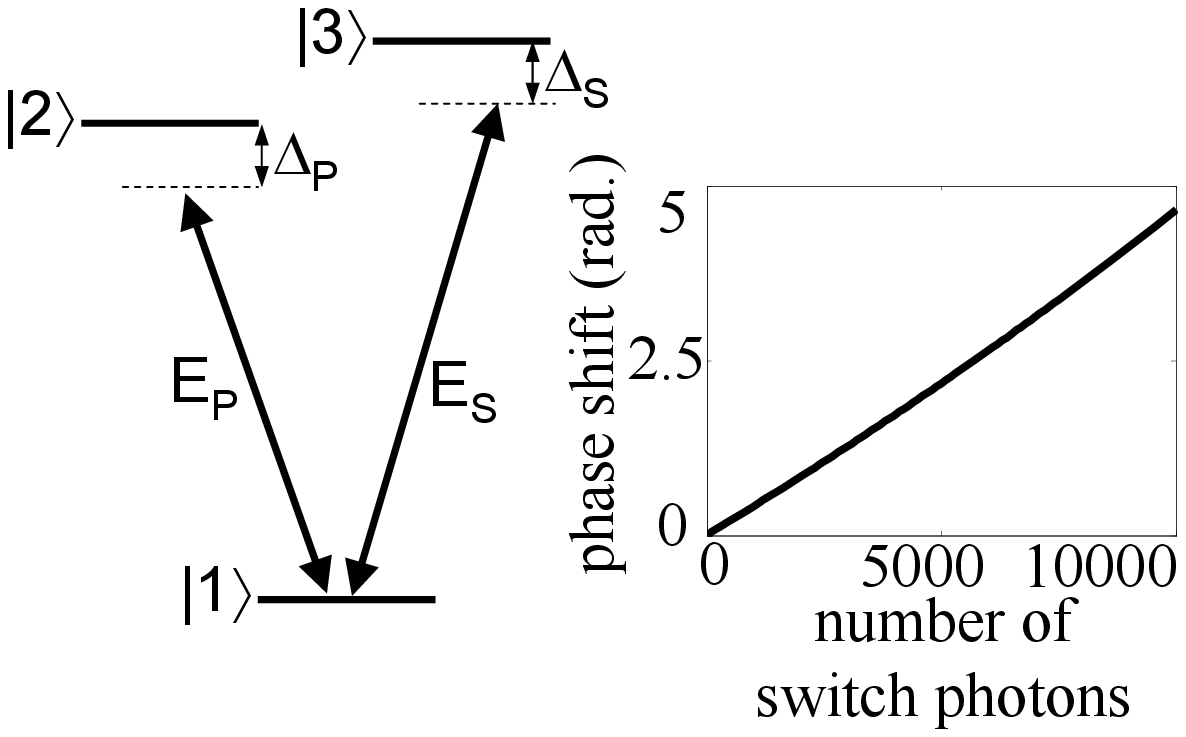}
\end{center}
\caption{A switch beam, $E_s$, causes a nonlinear phase shift on a probe beam, $E_p$.  The two beams travel collinearly through a V-type atomic medium.  By itself, the probe accumulates phase based on the linear susceptibility of the atoms.  When the switch beam is turned on, the common ground state $|1\rangle$ experiences an ac Stark shift, changing the effective detuning.  The plot indicates that $\sim5000$ switch photons are required for a phase shift of $\pi$ radians on the probe.} \label{levels}
\end{figure}

\begin{eqnarray}
\chi_p = \frac{N\mu_{12}^{2}}{\hbar\epsilon_0}\frac{1}{
2\left(\Delta_p+\delta_s\right)
+i\Gamma},
\label{SuscEqn}
\end{eqnarray}
where $\delta_s = \frac{\Omega_s^{2}}{4\Delta_s}$ is the ac Stark shift of the ground state, $\Omega_s$ is the Rabi frequency of the switch beam, $N$ is the atomic density, $\mu_{12}$ is the dipole matrix element between states $|1\rangle$ and $|2\rangle$, and $\Gamma$ is the transition linewidth. In the perturbative limit where $\delta_s\ll\Delta_p$, the nonlinear interaction between the switch and the probe can be described with a third order $\chi^{(3)}$ susceptibility by expanding Eq.~(\ref{SuscEqn}).  The polarization of the atomic medium at the probe laser frequency is then $P_p=\epsilon_0\chi^{(1)}E_p+\epsilon_0\chi^{(3)}E_s^{*}E_sE_p$.  In the ideal case of pure radiative broadening of the excited states, and in the limit where the detunings are much larger than the linewidth ($\Delta_p, \Delta_s \gg \Gamma$), the conditional phase shift (CPS) and absorption of the probe beam is:
\begin{eqnarray}
\text{CPS} \simeq n_s \left(\frac{3}{8\pi}\right)
\left(\frac{\lambda_s^{2}}{A}\right)
\left(\frac{1}{\tau\Delta_s}\right)
\left(\frac{\Gamma}{\Delta_p}\right) \phi^{(1)} \nonumber
\\
\text{absorption} \simeq
\left(\frac{\Gamma}{\Delta_p}\right)\phi^{(1)}.
\label{RealRatio}
\end{eqnarray}
Here, $n_s$ is the number of photons in the switch pulse and $\phi^{(1)}=\left(\frac{\omega_p}{2c}\right)N\Re(\chi^{(1)})L$ is the usual probe phase accumulation in the absence of the switch beam ($L$ is the length of the medium). In Eqs. \ref{RealRatio}, $\lambda_s$ is the wavelength of the switch field, $A$ is the spatial cross-sectional area of the two beams, and $\tau$ is the pulse duration of the two beams (for simplicity we take the two
beams to have the same temporal profile and assume them to be focused to the same size). To avoid significant reshaping of the beams, we must choose the bandwidth of the beams to be much smaller when compared with the detunings, $1/\tau \ll \Delta_p, \Delta_s$. From Eqs.~\ref{RealRatio}, for a high transmission of $> 50 \%$ and for the ideal case of $A \sim \lambda^2$, a CPS of $\pi$ radians requires a few thousand photons.

The plot in Fig.~\ref{levels} shows a numerical example based on Eq.~(\ref{SuscEqn}).  Here we use parameters that are typical for alkali atoms: wavelength $\lambda =780 $~nm and decay rate $\Gamma = 2 \pi \times 6$~MHz. We take $\Delta_p= \Delta_s = 160 \Gamma$, $NL = 1.5\times10^{13}$~cm$^{-2}$, $\tau = 20$~ns, and assume the ideal case of $A = \lambda^2$. With such tight focusing, propagation over long distances can, for example, be accomplished inside a hollow photonic crystal fiber \cite{Gaeta}. For these parameters, the transmission of the probe beam at the end of the medium is $60$~\%. We find a CPS of $\pi$ radians on the probe beam for a switch pulse containing 5000 photons. As we will discuss below, exact numerical calculations verify these results and demonstrate insignificant reshaping of the beams while propagating through the medium.

\begin{figure}[tb]
\begin{center}
\includegraphics[width=8.3cm]{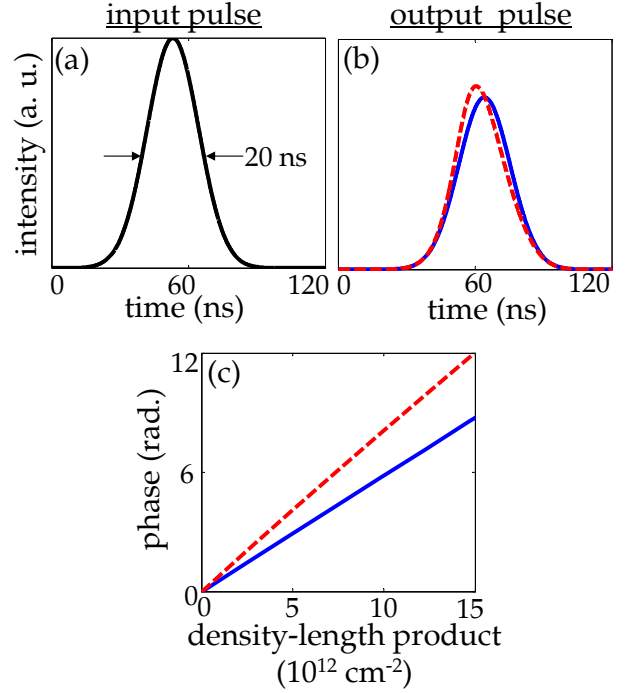}
\end{center}
\caption{(Color online) Numerical simulation of probe beam interacting with
switch beam in an atomic medium.  (a) shows the input probe pulse.
(b) shows the resultant probe pulse in the case that the switch is
off (blue, solid line) and on (red, dashed line).  (c) shows  the
probe pulse phase accumulation as a function of distance both with
the switch off (blue, solid line) and with the switch on (red,
dashed line).  The switch pulse (not shown here) has a matching
pulse-shape and frequency detuning.  This means the pulses stay
matched throughout the interaction.} \label{numericalFig}
\end{figure}

We proceed with a numerical study of the system. We neglect Doppler broadening and collisional effects and begin with a Hamiltonian describing a closed, three-level V-system in local time $t'=t-z/c$
\begin{eqnarray}
\mathcal{H} = \hbar\begin{pmatrix}
  0 & -\frac{\Omega_p(z,t')}{2} & -\frac{\Omega_s(z,t')}{2} \\
  -\frac{\Omega_p(z,t')^*}{2} & \Delta_p & 0 \\
  -\frac{\Omega_s(z,t')^*}{2} & 0 & \Delta_s \\
\end{pmatrix}.
\label{Hamiltonian}
\end{eqnarray}
We then use the commutator and anticommutator relations to find the equation of motion for the three-by-three density matrix $\rho$ \cite{ScullyBook}:
\begin{eqnarray}
\dot{\rho} = -\frac{i}{\hbar}[\mathcal{H},\rho]-\frac{1}{2}\{\Gamma,\rho\}.
\label{EqnOfMotion}
\end{eqnarray}
The values of $\rho_{ij}$ calculated in Eq.~(\ref{EqnOfMotion}) are used to numerically integrate the slowly varying envelope Maxwell's equations governing the propagation of the probe and switch fields,
\begin{eqnarray}
\frac{\partial \Omega_p(z,t')}{\partial
z}=-\frac{i}{\hbar}\eta\omega_p N\mu_{12}^2 \rho_{12}(z,t')
\nonumber
\\
\frac{\partial \Omega_s(z,t')}{\partial
z}=-\frac{i}{\hbar}\eta\omega_s N\mu_{13}^2 \rho_{13}(z,t'),
\label{Maxwell}
\end{eqnarray}
where $\eta=\sqrt{\mu_0/\epsilon_0}$ is the impedance of free space. We solve Eqs.~(\ref{EqnOfMotion}) and (\ref{Maxwell}) with the initial condition that all atoms are in ground state $|1\rangle$. At the start of the atomic medium ($z=0$) we apply a boundary condition that the fields, and therefore the Rabi frequencies $\Omega_p(z=0,t')$ and $\Omega_s(z=0,t')$, are long Gaussian envelopes with a Gaussian width of $\tau$. Eqs. (\ref{EqnOfMotion}) and (\ref{Maxwell}) are then solved on the space-time grid using the method of lines.

\begin{figure}[tb]
\begin{center}
\includegraphics[width=8.3cm]{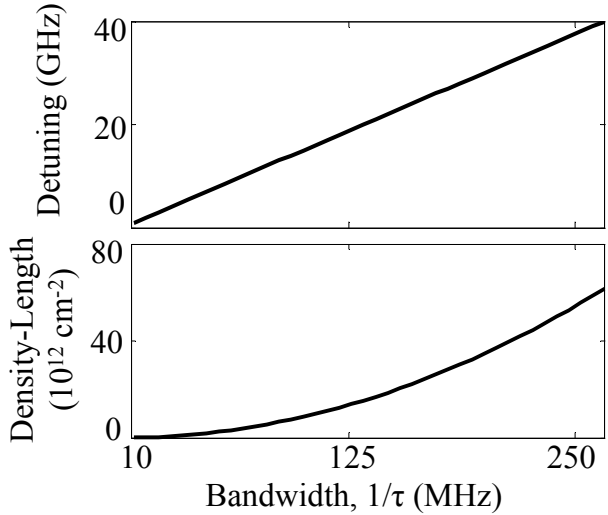}
\end{center}
\caption{The probe and switch pulses can be made shorter,
broadening the bandwidth, at the expense of density-length
product.  As the bandwidth is broadened, the detuning must
increase to avoid near-resonant effects.  If the density-length
product is appropriately increased, identical results in
transmission and phase-shift are obtained.  Since faster pulses
result in higher intensity for the same energy, the required
number of switch beam photons does not change.}
\label{bandwidthFig}
\end{figure}
The results are presented in Fig.~\ref{numericalFig} and demonstrate a phase-shift of 3.2 radians with 60\% transmission. In this simulation, we use the same parameters as the plot in Fig.~\ref{levels} and use $n_s=5000$ photons in the switch beam.  We observe smooth time-profiles at the end of the medium demonstrating negligible reshaping. Since the probe and the switch beam have identical detunings from the excited state, the switch pulse (not plotted) experiences similar absorption and reshaping.  Furthermore, the two beams propagate with the same group velocity and therefore stay spatially and temporally well-matched while propagating through the medium.

Finally, we note that the energy, or number of switch photons, required for this phase shifter is independent of bandwidth.  Fig.~\ref{bandwidthFig} shows the required detuning and the density-length product for a given bandwidth that achieves the same performance as the numerical simulation of Fig.~\ref{numericalFig} (a CPS of $\sim\pi$ radians for $n_s=5000$ switch photons). As the bandwidth broadens, both the probe and the switch must be appropriately detuned to avoid near-resonance effects.  As noted in Eq.~(\ref{RealRatio}), the increased switch detuning is compensated by the shortened pulse duration (increased bandwidth), which means the switch pulse is more intense for the same energy. The increased probe detuning trades off with an increased density-length product to keep the probe transmission constant. The density-lengths required for a fast ($>$100 MHz) phase shifter in a typical alkali atom have been achieved in cold atom traps and optical fibers containing rubidium vapor \cite{Thad,Gaeta}.

In summary, we suggested a far-off resonant scheme that supplies a conditional phase shift of $\pi$ radians with energies at the 1000 photon level. To the best of our knowledge, the phase shifter presented here is among the simplest of those suggested in the literature. As mentioned before, a possible application of our suggestion is to all-optical information processing. With our approach, it should be possible to construct an all-optical switch with a switching time approaching 1 nanosecond at a total energy cost of less than 1 femtojoule per switching event. Furthermore, by using a cavity of a finesse of about 1000, our approach may achieve switching at the single photon level. If the switch beam can be supplied by a single photon, then the suggestion described here may be applicable as a single-photon controlled-NOT gate between the probe and the switch.  This will be among our future investigations.

We thank J.~T.~Green for helpful discussions. This work was supported by Air Force Office of Scientific Research (AFOSR) and University of Wisconsin Alumni Research Foundation (WARF).


\end{document}